\documentclass[12pt,english]{article}
\pdfoutput=1
\usepackage[T1]{fontenc}
\usepackage{graphicx,array}
\usepackage{cite}
\usepackage{color}
\usepackage{bm}
\usepackage{dsfont}
\usepackage{latexsym}
\usepackage{amsthm}
\usepackage{amsmath}
\usepackage{stackengine}
\usepackage[titletoc]{appendix}
\usepackage{enumitem}
\usepackage{amssymb}
\usepackage{float}
\usepackage[unicode=true,
 bookmarks=true,bookmarksnumbered=false,bookmarksopen=false,
 breaklinks=false,pdfborder={0 0 1},backref=false,colorlinks=true]
 {hyperref}
\hypersetup{pdftitle={Large Diffeomorphisms and Accidental Symmetry of the Extremal Horizon},
 pdfauthor={Achilleas P. Porfyriadis, Grant N. Remmen},
 citecolor=blue,linkcolor=black,urlcolor=black}
\usepackage{breakurl}
\usepackage[hang,flushmargin]{footmisc}

\newcolumntype{L}[1]{>{\raggedright\let\newline\\\arraybackslash\hspace{0pt}}m{#1}}
\newcolumntype{C}[1]{>{\centering\let\newline\\\arraybackslash\hspace{0pt}}m{#1}}

\def\simgt{\mathrel{\lower2.5pt\vbox{\lineskip=0pt\baselineskip=0pt
           \hbox{$>$}\hbox{$\sim$}}}}
\def\simlt{\mathrel{\lower2.5pt\vbox{\lineskip=0pt\baselineskip=0pt
           \hbox{$<$}\hbox{$\sim$}}}}

\newcommand{\be}{\begin{equation}}
\newcommand{\ee}{\end{equation}}
\newcommand{\bea}{\begin{eqnarray}}
\newcommand{\eea}{\end{eqnarray}}
\newcommand{\Eq}[1]{Eq.~(\ref{#1})}
\newcommand{\Sec}[1]{Sec.~\ref{#1}}

\newcommand*\oline[1]{%
  \vbox{%
    \hrule height 0.5pt
    \kern0.68ex
    \hbox{%
      \kern-0.1em
      \ifmmode#1\else\ensuremath{#1}\fi
      \kern-0.1em
    }
  }
}

\definecolor{nicered}{rgb}{0.7,0.1,0.1}
\definecolor{nicegreen}{rgb}{0.1,0.5,0.1}
\hypersetup{
 colorlinks,citecolor=black,linkcolor=black,urlcolor=black}
\usepackage{xcolor}

\renewcommand{\Box}{\square}

\begin{document}

\interfootnotelinepenalty=10000
\baselineskip=18pt
\hfill

\vspace{2cm}
\thispagestyle{empty}
\begin{center}
{\LARGE \bf
Large Diffeomorphisms and Accidental Symmetry of the Extremal Horizon
}\\
\bigskip\vspace{1cm}{
{\large Achilleas P. Porfyriadis${}^{a,b}$ and Grant N. Remmen${}^{c,d}$}
} \\[7mm]
{
\it ${}^a$Center for the Fundamental Laws of Nature, Harvard University, Cambridge, MA 02138
${}^b$Black Hole Initiative, Harvard University, Cambridge, MA 02138\\[1.5 mm]
${}^c$Kavli Institute for Theoretical Physics, University of California, Santa Barbara, CA 93106
${}^d$Department of Physics, University of California, Santa Barbara, CA 93106 \\[1.5 mm]}
\let\thefootnote\relax\footnote{e-mail: 
\url{porfyr@g.harvard.edu}, \url{remmen@kitp.ucsb.edu}}
 \end{center}

\bigskip
\centerline{\large\bf Abstract}
We uncover a symmetry of the linear Einstein equations near extremal horizons. 
Specifically, acting with a spherically symmetric linearized diffeomorphism
on the perturbative solutions to the Einstein-Maxwell equations in the Bertotti-Robinson background, but not acting on the background itself, we find that there is a subset of such transformations under which the equations of motion remain satisfied, with or without additional matter.
This represents 
an ``accidental'' symmetry in the sense that the set of transformations realizing the mapping among solutions is strictly larger than the ${\rm SL}(2)$ isometries of the background spacetime.
We argue that our accidental symmetry can be thought of as an \emph{on-shell} large diffeomorphism of ${\rm AdS}_2$, which we support in the context of Jackiw-Teitelboim theory.

\begin{quote} \small

\end{quote}
	
\setcounter{footnote}{0}

\newpage
\tableofcontents
\newpage

\section{Introduction}\label{sec:intro}

Coordinate transformations may sometimes be more than mere reparameterizations of the same physical reality and under certain conditions map one physical state to another. 
In gravitational theories, such coordinate transformations are typically associated with the presence of a boundary that prevents certain diffeomorphisms from being trivial reparameterizations by imposing on them nontrivial boundary conditions. 
For example, in four-dimensional asymptotically flat general relativity, boundary conditions at null infinity render physical the so-called supertranslations $u\to u+f(z,\bar{z})$, which shift the retarded time $u$ at null infinity by an arbitrary function $f$ on the celestial two-sphere \cite{Bondi:1962px,Sachs:1962wk}. Similarly, on the timelike boundary in three-dimensional asymptotically anti-de Sitter (AdS) space, arbitrary holomorphic transformations $z\to f(z), \bar{z}\to \bar{f}(\bar{z})$ of the boundary coordinates map one physical state in gravity to another because of the boundary conditions imposed on the corresponding bulk diffeomorphisms \cite{Brown:1986nw}.
When diffeomorphisms become physical, they manifest themselves at both quantum and classical level. For example, the supertranslations have been identified both with Weinberg's soft graviton theorem and with the gravitational memory effect~\cite{Strominger:2017zoo}, and the two-dimensional local conformal transformations are at play in a vast series of results in ${\rm AdS}_3/{\rm CFT}_2$ holography. 
Because they are symmetries, it does not come as a surprise that physical diffeomorphisms arise in a multitude of ways. However, it is curious that when such physical maps are discovered in gravitational analyses, they often remain largely unused for a long time. For example, the symmetries of both null infinity in four dimensions and the timelike boundary of ${\rm AdS}_3$ were discovered in gravitational studies of asymptotic symmetry groups long before the present surge in investigations of celestial holography and the discovery of AdS/CFT, respectively.\footnote{Bondi, van der Burg, Metzner, and Sachs (BMS) found the supertranslations in 1962, and Brown-Henneaux computed the eponymous central charge in 1986.}
In the case of ${\rm AdS}_2$, it has recently been realized from the holographic perspective that here, too, its boundary time reparametrizations $t\to f(t)$ are associated with bulk diffeomorphisms that are physical~\cite{Jensen:2016pah,Maldacena:2016upp,Engelsoy:2016xyb}, while a purely gravitational asymptotic symmetry group argument already existed in Refs.~\cite{Hotta:1998iq,Cadoni:1999ja,Navarro-Salas:1999zer}.

Asymptotic symmetry group analyses rely crucially on the existence of a boundary in the spacetime with respect to which boundary conditions are imposed and asymptotic expansions away from which are used to define the set of diffeomorphisms that will end up being physical. Indeed, in this case the physical diffeomorphisms are often called ``large,'' as they do not fall off fast enough to be trivial near the boundary. Large diffeomorphisms may be usefully thought of as deforming the shape of the boundary and therefore changing the boundary data that are used to define the gravitational theory in the interior. This raises the question: What if the boundary is only an approximate notion, as in the case with ${\rm AdS}_2$ when it arises in the near-horizon approximation of (near-)extreme black holes? Because this ${\rm AdS}_2$ is only approximate, it does not come with its timelike boundary, which in this case is replaced by a matching region where the asymptotically flat part of the black hole geometry is glued onto the ${\rm AdS}_2$ part. Under such circumstances, are there still diffeomorphisms associated with ${\rm AdS}_2$ that are physical in an appropriate sense? The purpose of this paper is to answer this question in the affirmative. 

To leading order in the deviation from extremality, both an extreme and a near-extreme black hole have a near-horizon geometry given by ${\rm AdS}_2$, albeit in different coordinate patches (the Poincar\'e and Rindler patches, respectively). 
In Ref.~\cite{Hadar:2020kry}, it was explained in detail that the distinction between the two black hole spacetimes may be detected in the linear perturbations away from the ${\rm AdS}_2$ geometry. Indeed, for the spherically symmetric electrovacuum Einstein equations, the nontrivial perturbations of ${\rm AdS}_2\times S^2$ are a three-parameter family of solutions.
A quadratic of the three parameters, dubbed $\mu$, may be identified with the Casimir of the ${\rm SL}(2)$ isometry group of ${\rm AdS}_2$ and determines whether the back reaction of the perturbations on the ${\rm AdS}_2\times S^2$ background leads to an extreme ($\mu=0$) or near-extreme ($\mu >0$) Reissner-Nordstr\"om (RN)  black hole. This back reaction that changes the ${\rm AdS}_2$ asymptotics to produce an asymptotically flat spacetime was called anabasis in Ref.~\cite{Hadar:2020kry}. 

The ${\rm SL}(2)$ isometries of the background are guaranteed to map one perturbative solution to another, but they do not change the quadratic $\mu$, which is ${\rm SL}(2)$ invariant. On the other hand, acting with arbitrary diffeomorphisms on the perturbations alone, while leaving the background untouched, one has no reason to expect that the result will still produce solutions to the linear Einstein equations around the original background. 
Remarkably, we show that there in fact exists a one-parameter family of such transformations that map one perturbative solution with an ${\rm SL}(2)$ invariant $\mu$ to another one with a different invariant $\mu'$. Such diffeomorphisms deserve to be called physical because they form a map between solutions that, upon back reaction, give rise to black holes with different charge-to-mass ratios (i.e., different deviations from extremality). 
Borrowing terminology from particle physics, where a symmetry is described as ``accidental'' if it happens to be respected by the low-energy action without being protected in the fundamental theory, our symmetries can similarly be called ``accidental,'' since they describe transformations among the perturbative solutions despite not being isometries of the background itself.
For example, consider taking an extreme RN black hole with its ${\rm AdS}_2\times S^2$ near-horizon geometry (in Poincar\'e coordinates) and its linear perturbation, for which $\mu=0$. This perturbation, by definition, is such that when it back reacts on ${\rm AdS}_2\times S^2$, it leads to anabasis and builds the full extreme black hole. Acting on it with one of our physical diffeomorphisms produces another linear solution, this time with $\mu>0$, and if we let it back react on the same ${\rm AdS}_2\times S^2$, it now leads to a near-extreme black hole with a deviation from extremality measured by the new $\mu$. 

This phenomenon takes its starkest form when there are also propagating degrees of freedom supported by the equations. To this end, still in the spherically symmetric sector, we also consider the Einstein equations with a minimally coupled scalar that solves the wave equation on ${\rm AdS}_2$. In this case, we show that there exist physical diffeomorphisms that map the empty black hole solution to one with scalar waves turned on. Specifically, 
we show that for an arbitrary spherically symmetric scalar $\phi\sim\mathcal{O}(\lambda^{1/2})$ that solves the Klein-Gordon equation on ${\rm AdS}_2\times S^2$, the corresponding linear perturbation $h_{\mu\nu}\sim\mathcal{O}(\lambda)$ is related to the $\mu=0$ anabasis perturbation of the empty extreme black hole by an appropriate physical diffeomorphism. In other words, starting with a static extreme black hole, there exists a physical diffeomorphism that may be used to turn on any spherically symmetric linear propagating wave degree of freedom! It should be noted that this linear perturbation $h_{\mu\nu}$, upon back reaction, generically produces a near-extreme black hole with $\mu=\mathcal{O}(\lambda)$, as expected~\cite{Murata:2013daa}. As a result, we conclude that in the spherically symmetric sector of the Einstein-Maxwell-scalar equations of motion, all solutions near the empty extreme RN black hole may be obtained from it by an appropriate physical diffeomorphism applied on the linear anabasis perturbation of its ${\rm AdS}_2\times S^2$ throat.

The physical diffeomorphisms in this paper may be thought of as ``on-shell large diffeomorphisms'' in the following sense. Ordinary large diffeomorphisms, derived as members of an asymptotic symmetry group, change at will the data prescribed on the boundary with respect to which they are defined. In the ${\rm AdS}_2$ case, this means that the large diffeomorphisms $t\to f(t)$ for arbitrary $f$ are all physical. Equivalently, one may view these large diffeomorphisms as arbitrarily deforming the shape of the ${\rm AdS}_2$ boundary, thereby defining a physical state in the interior for arbitrary boundary shapes. In AdS/CFT, one says that one may prescribe arbitrarily the sources and vacuum expectation values (vevs) of the fields in the boundary CFT, and this defines the gravitational theory in the AdS bulk.
In our case, with the ${\rm AdS}_2$ embedded in an asymptotically flat spacetime, if we imagine that the boundary of ${\rm AdS}_2$ is replaced by a timelike matching surface, then the Einstein equations across this surface impose constraints on its allowed shape deformations. This reduces the set of physical diffeomorphisms from all large ${\rm AdS}_2$ diffeomorphisms to those that are also consistent with the Einstein equations in the entire asymptotically flat spacetime. In other words, in our setup, we are classically putting the large diffeomorphisms of ${\rm AdS}_2$ on shell. In the language of AdS/CFT, the connection of the ${\rm AdS}_2$ with the asymptotically flat region provides an independent relation between the sources and vevs on the matching surface, a relation that is dictated by the Einstein equations in the asymptotically flat region.
We will discuss further the on-shell condition in Sec.~\ref{Sec:Discussion} from the point of view of the Schwarzian action for ${\rm AdS}_2$ gravity  in the Jackiw-Teitelboim (JT) theory.

This paper is structured as follows. Sec.~\ref{sec:pertns} contains a review of the spherically symmetric electrovacuum perturbations of ${\rm AdS}_2\times S^2$ and their back reaction properties in terms of building RN black holes at or near extremality. In Sec.~\ref{sec:symsec}, we define our notions of physical diffeomorphisms and accidental symmetries in the linear Einstein equations, and we solve for them in the electrovacuum case near the horizon of extreme RN. Sec.~\ref{sec:mattersec} generalizes the results with the addition of propagating matter degrees of freedom. The discussion in Sec.~\ref{Sec:Discussion} compares and contrasts our physical diffeomorphisms with on-shell and off-shell large diffeomorpisms of ${\rm AdS}_2$ in the context of JT theory.

\section{Perturbations near the horizon}\label{sec:pertns}

The near-horizon geometry of an extremal RN black hole famously takes the form of the direct product of ${\rm AdS}_2$ with an $S^2$, also known as the  Bertotti-Robinson solution.
We will write this geometry in Poincar\'e-AdS coordinates,
\be
{\rm d}s^2 =  -r^2 {\rm d}t^2 + \frac{{\rm d}r^2}{r^2} + {\rm d}\Omega^2,\label{eq:background}
\ee
with the gauge field describing a magnetic black hole,
\be
F =  \sin \theta\,{\rm d\theta}\wedge {\rm d \varphi}.
\ee
In these coordinates, the horizon is located at $r=0$.
We work in units with $8\pi G = 1$, and we also set the common length scale of the ${\rm AdS}_2$ and $S^2$ factors ($\propto \sqrt{G}\, q$ for a black hole with charge $q$) to unity throughout, since it can be straightforwardly restored by dimensional analysis.

Let us now perturb this near-horizon geometry, sending the metric to $g_{\mu\nu} + \delta g_{\mu\nu}$ and the gauge field strength to $F_{\mu\nu} + \delta F_{\mu\nu}$. Expanding the background Einstein-Maxwell equations,
\be
R_{\mu\nu} - \frac{1}{2}Rg_{\mu\nu} = F_{\mu\rho}F_\nu^{\;\;\rho} -\frac{1}{4}g_{\mu\nu}F_{\rho\sigma}F^{\rho\sigma} \label{eq:ein0}
\ee
and 
\be
\nabla_\mu F^{\mu\nu} = 0, \label{eq:max0}
\ee
to first order in $\delta$, we obtain the linearized equations $\delta E_{\mu}^{\;\;\nu} = 0$ and $\delta M^\mu = 0$, respectively, where for convenience we raise an index on the Einstein equation.
We find that the linearized equations simplify if we consider the equivalent expressions obtained by subtracting traces or the background equations of motion.
Writing Eqs.~\eqref{eq:ein0} and \eqref{eq:max0} as $E_\mu^{\;\;\nu} = 0$ and $M^\mu = 0$, respectively, we consider $\delta \hat E_\mu^{\;\;\nu} =\delta E_\mu^{\;\;\nu} -\frac{1}{2}\delta_\mu^\nu \delta E_\rho^{\;\;\rho}+  \left(E_\mu^{\;\;\rho} -\frac{1}{2}\delta_\mu^\rho E_\sigma^{\;\;\sigma}\right)\delta g_\rho^{\;\;\nu}$ and $\delta \hat M^\mu = \delta M^\mu + M^\nu \delta g_\nu^{\;\;\mu}$.
For convenience writing $\delta g_{\mu\nu} = h_{\mu\nu}$ and $\delta F_{\mu\nu} = f_{\mu\nu}$, we have the linearized Einstein-Maxwell equations of motion,
\be
\begin{aligned}
\delta \hat E_{\mu}^{\;\;\nu} & =\frac{1}{2}\left(\nabla_{\rho}\nabla_{\mu}h^{\rho\nu}+\nabla^{\rho}\nabla^{\nu}h_{\rho\mu}-\Box h_{\mu}^{\;\;\nu}-\nabla_{\mu}\nabla^{\nu}h_{\rho}^{\;\;\rho}\right)\\
 & \qquad-\delta_{\mu}^{\nu}F_{\rho\alpha}F_{\sigma}^{\;\;\alpha}h^{\rho\sigma}+2F_{\mu\rho}F^{\nu\sigma}h_{\;\;\sigma}^{\rho}+\frac{1}{2}F_{\rho\sigma}F^{\rho\sigma}h_{\mu}^{\;\;\nu}-2F_{\mu\rho}f^{\nu\rho}-2F^{\nu\rho}f_{\mu\rho}+\delta_{\mu}^{\nu}F^{\rho\sigma}f_{\rho\sigma}
\end{aligned}\label{eq:ein}
\ee
and
\be
\delta \hat M^{\mu}  =\nabla_{\nu}f^{\nu\mu}-h^{\nu\rho}\nabla_{\rho}F_{\nu}^{\;\;\mu}-F_{\rho}^{\;\;\mu}\nabla_{\nu}h^{\nu\rho}-F^{\nu\rho}\nabla_{\nu}h_{\rho}^{\;\;\mu}+\frac{1}{2}F^{\nu\mu}\nabla_{\nu}h_{\rho}^{\;\;\rho}.\label{eq:max}
\ee
We consider spherically symmetric perturbations,
\be
\begin{aligned}
h_{\mu\nu}{\rm d}x^\mu {\rm d}x^\nu &= r^2 H_0(t,r) \,{\rm d}t^2 + 2 H_1 (t,r)\,{\rm d}t\,{\rm d}r + \frac{H_2(t,r)}{r^2}{\rm d}r^2 + K(t,r){\rm d}\Omega^2\\
f_{\mu\nu}\,{\rm d}x^\mu \wedge {\rm d}x^\nu &= u(t,r) \sin\theta\,{\rm d}\theta\wedge {\rm d}\varphi.
\end{aligned}
\ee
Without loss of generality, we fix our gauge to be radial, the so-called Fefferman-Graham (FG) gauge for AdS, in which $H_1(t,r) = H_2(t,r) = 0$.
The general solution to the linearized Einstein-Maxwell equations in FG gauge is then given by
\be
\begin{aligned}
K(t,r) &=  \Phi_0 + \Phi \\
H_0(t,r) &= \Phi_0 \log r  +2 a r+2 b r t + 2 c r\left(t^2 + \frac{3}{r^2}\right)+ \frac{C_1'''(t)}{r^2} +2 C_2'(t) \\
u(t,r) &= \Phi_0.
\end{aligned}\label{eq:HKu}
\ee
Here, 
\be \label{Phi}
\Phi = a r + b r t + c r\left(t^2 - \frac{1}{r^2} \right),
\ee
with $(a,b,c,\Phi_0)$ arbitrary constants and $C_{1,2}(t)$ arbitrary functions. The latter are pure gauge and may be set to zero by using the residual gauge freedom left in keeping the FG condition satisfied, namely by adding the pure gauge perturbation $2\nabla_{(\mu} \chi_{\nu)}$ to $\delta g_{\mu\nu}$, for $\chi = \left[C_1(t)+C_2(t) + \frac{C_1''(t)}{2r^2}\right]\partial_t -r C_1'(t)\partial_r$.
The function $K(t,r) = h_{\theta\theta}$ is invariant under spherically symmetric gauge transformations in the ${\rm AdS}_2 \times S^2$ background.

The unperturbed ${\rm AdS}_2 \times S^2$ geometry in \Eq{eq:background} is invariant under time translations $t\rightarrow t+\alpha$, dilations $(t,r)\rightarrow \left({t}/{(1+\beta)},(1+\beta)r\right)$, and special conformal transformations (SCT), which may be written as finite coordinate transformations $x^\mu \rightarrow x^\mu + \Delta x^\mu_{0,1,2}$, with $\Delta x^\mu_{0,1,2}$ taking the form
\be 
\begin{aligned}
\Delta x_0(\alpha) &= \alpha\partial_t & [\text{time translation}] \\
\Delta x_1(\beta) &= -\frac{\beta}{1+\beta}t\partial_t +  \beta r\partial_r & [\text{dilation}]  \\
\Delta x_2(\gamma) &= \left[\frac{t-\gamma(t^2 - r^{-2})}{1-2\gamma t  + \gamma^2 (t^2 - r^{-2})} - t\right]\partial_t - r\left[2\gamma t - \gamma^2(t^2 - r^{-2})\right]\partial_r. &[\text{SCT}]
\end{aligned}
\ee
As discussed in Ref.~\cite{Hadar:2020kry}, the linearized solutions to the equation of motion can be naturally categorized by their behavior under ${\rm SL}(2)$ transformations of the gauge-invariant term $K(t,r) = \Phi_0 + \Phi$.
Let us briefly review this categorization.
The $\Phi_0$ solution transforms, within $K$, as a singlet under ${\rm SL}(2)$.
The $(a,b,c)$ solutions encoded in $\Phi$, however, break the ${\rm SL}(2)$ symmetry and form an ${\rm SL}(2)$ triplet, which transforms as
\be
\begin{aligned}
\Delta x^\mu_0(\alpha):\qquad &&  \Delta(a,b,c) &= (\alpha b + \alpha^2 c, 2\alpha c,0) \\
\Delta x^\mu_1(\beta):\qquad && \Delta(a,b,c) &= \left(\beta a,0,-\frac{\beta}{1+\beta }c\right) \\
\Delta x^\mu_2(\gamma):\qquad && \Delta(a,b,c) &= (0,-2\gamma a, \gamma^2 a- \gamma b ).
\end{aligned}\label{eq:abctransform}
\ee
Thus, the $a$ solution preserves only time translations, the $b$ solution only dilations, and the $c$ solution only SCT.
The combination 
\be
\mu = b^2 - 4 a c 
\ee
is the quadratic Casimir of ${\rm SL}(2)$ and, as one can verify, is invariant under the transformations in \Eq{eq:abctransform}.

The parameter $\mu$ is useful in distinguishing perturbative geometries. Solutions in the $(a,b,c)$ triplet with a given, fixed $\mu$ can be mapped into each other by following the orbit of ${\rm SL}(2)$.
On the other hand, solutions with different $\mu$ are inequivalent, even after modding out by ${\rm SL}(2)$ transformations.
In fact, $\mu$ has physical significance in the context of charged black holes.
Consider the RN metric,
\be
{\rm d}s^2 = -f(\hat t) {\rm  d}\hat t^2 + \frac{{\rm d}\hat r}{f(\hat r)} + \hat r^2 {\rm d}\Omega^2, 
\ee
where $f(\hat r) = 1 - \frac{2M}{\hat{r}} + \frac{Q^2}{\hat{r}^2}$. 
Taking the extremal case where $Q=M$, defining new coordinates (in $M=1$ units),
\be
\begin{aligned}
\hat t &= t/\lambda \\
\hat r &= 1 + \lambda r - \lambda^2 r^2,
\end{aligned} 
\ee
and taking the near-horizon limit of $\lambda\rightarrow 0$ produces the Bertotti-Robinson metric of \Eq{eq:background}, with the ${\cal O}(\lambda)$ perturbation given in FG gauge by \Eq{eq:HKu}, with $a = 2\lambda$ and $b = c = \Phi_0 = 0$.
Thus, the perturbations to ${\rm AdS}_2 \times S^2$ that correspond to the near-horizon throat of an exactly extremal RN black hole have vanishing ${\rm SL}(2)$ invariant $\mu = 0$.
In contrast, suppose we instead consider a subextremal black hole, with $Q/M = \sqrt{1-\lambda^2 \kappa^2}$ for some $\kappa > 0$. Defining (in $M=1$ units), 
\be
\begin{aligned}
\hat t &= -\frac{1}{2\lambda \kappa}\log\left|t^2 - \frac{1}{r^2}\right| + 1 + \frac{1}{1-rt} - \frac{1}{1+rt} - 2\, {\rm arctanh}\,rt \\
\hat r &= (1+\lambda\kappa)[1-\lambda\kappa(1+rt)]+\lambda^2 \kappa^2 rt(1-rt),
\end{aligned}
\ee
one finds that in the near-horizon ($\lambda \rightarrow 0$) limit, the geometry takes the form of \Eq{eq:background}, with the leading ${\cal O}(\lambda)$ perturbation again given in FG gauge as in \Eq{eq:HKu}, but with $b = -2\lambda\kappa$ and $a=c=\Phi_0 = 0$.
That is, the near-extremal RN black hole possesses a near-horizon ${\rm AdS}_2 \times S^2$ geometry, with the first correction away from the horizon solving the linearized Einstein-Maxwell equations in the ${\rm AdS}_2 \times S^2$ background, corresponding to an ${\rm SL}(2)$ invariant $\mu > 0$.

\section{Symmetries of the extreme-horizon Einstein equations}\label{sec:symsec}

We now examine the linearized Einstein equations near the extreme horizon in more detail and uncover a surprising enhanced symmetry in their structure.

\subsection{Diffeomorphisms and enhanced symmetry}\label{sec:enhancedsymmetry}

We can view the linearized Einstein equation~\eqref{eq:ein} in a background geometry $\bar g$ as a linear differential operator ${\cal E}(\bar g,h)$ acting on some metric perturbation $h_{\mu\nu}$ for which perturbative solutions satisfy ${\cal E}(\bar g,h) = 0$.\footnote{We could develop similar formalism for the Maxwell equations, but since we are considering a magnetic background, these will be trivially satisfied for the spherically symmetric perturbations we consider. Similarly, we will take $\Phi_0 = 0$ henceforth, so that the perturbation of the gauge field vanishes.}
We can act with a finite diffeomorphism, transforming the $(t,r)$ coordinates via
\be\label{transformation}
(t,r) \rightarrow (t,r) + \lambda (\xi^t(t,r),\xi^r(t,r)), 
\ee
under which both the background and perturbation transform as $\bar g \rightarrow \bar g(\lambda)$ and $h\rightarrow h(\lambda)$, where $\bar g = \bar g_{\mu\nu} {\rm d}x^\mu {\rm d}x^\nu$ and $h = h_{\mu\nu} {\rm d}x^\mu {\rm d}x^\nu$.
In particular, let us take our original background $\bar g$ to be that of the near-horizon extremal black hole, in Poincar\'e-AdS coordinates~\eqref{eq:background}.
By general covariance, for arbitrary $\lambda$ and $\xi^\mu$ the equations of motion remain satisfied,
\be 
{\cal E}(\bar g(\lambda),h(\lambda)) = 0.\label{eq:eomE}
\ee
Thus, \Eq{eq:eomE} must be true at each order in an expansion in $\lambda$.
To ${\cal O}(\lambda)$, we have
\be
{\cal E}(\bar{g}(0),h(0))+\lambda\frac{\delta}{\delta\lambda}{\cal E}(\bar{g}(\lambda),h(0))+\lambda\frac{\delta}{\delta\lambda}{\cal E}(\bar{g}(0),h(\lambda))+{\cal O}(\lambda^{2})=0 .
\ee
But ${\cal E}(\bar g(0),h(0))$ vanishes by definition, since we have started with a solution to the linearized Einstein equations around the original background.
Hence,
\be 
\lim_{\lambda\rightarrow 0}\left[\partial_\lambda{\cal E}(\bar{g}(\lambda),h(0))+\partial_\lambda{\cal E}(\bar{g}(0),h(\lambda))\right] =0.\label{eq:eomElinear}
\ee
Let us consider what these two terms describe. The first term holds the perturbation $h$ fixed, while acting with a linearized diffeomorphism on the background metric. 
The second term instead considers a fixed background and transforms the perturbation $h$ itself by acting with a linearized diffeomorphism.
Their effects must be equal and opposite.

However, we can impose a further, strong requirement on $\xi^\mu$.
Suppose that each of the two terms in \Eq{eq:eomElinear} vanish individually,
\be
\lim_{\lambda\rightarrow 0}\partial_\lambda {\cal E}(\bar{g}(0),h(\lambda)) = 0.\label{eq:eomEfinal}
\ee
In particular, this means that the Einstein equations around the fixed ${\rm AdS}_2\times S^2$ background $\bar g$ enjoy an enhanced symmetry, namely, their solutions---the various $h_{\mu\nu}$---transform among themselves under the action of $\xi^\mu$.
This is a strong requirement on the $\xi^\mu$, which is now no longer strictly a diffeomorphism, but is instead a physical, dynamically constrained symmetry of the equations of motion.
In other words, for a given solution with ${\rm SL}(2)$ parameters $(a_0,b_0,c_0)$ in the $(a,b,c)$ parameter space, described in FG gauge by \Eq{eq:HKu}, there is some map of the form $h\rightarrow h(\lambda)$ that connects any two $(a,b,c)$ solutions, modulo gauge, in some small neighborhood ${\cal N}$ of $(a_0, b_0,c_0)$.
The existence of this map is highly nontrivial.
Of course, among the solutions of fixed $\mu$ invariant $\mu_0 = b_0^2 - 4a_0 c_0$ within this neighborhood, this map is simply described by the ${\rm SL}(2)$ generators that form an isometry of the background itself, and as a result the first term in \Eq{eq:eomElinear} automatically vanishes, trivially leading to \Eq{eq:eomEfinal}.
But, crucially, ${\cal N}$ contains $(a,b,c)$ with $\mu\neq \mu_0$,  and the map given by physical diffeomorphisms fills out the full open ball around $(a_0,b_0,c_0)$ by connecting subspaces of ${\cal N}$ corresponding to different ${\rm SL}(2)$ orbits.
The map among solutions of differing $\mu$ is not provided by the ${\rm SL}(2)$ symmetry of the background geometry itself.
Its existence is surprising and indeed represents an accidental symmetry of the linear equations of motion.

\subsection{Solving for the map}
Consider a linearized solution as given in \Eq{eq:HKu}.
For $h_{\mu\nu}$ to describe the near-horizon geometry of an extremal black hole---i.e., deviations from strict ${\rm AdS}_2 \times S^2$ as we traverse the black hole throat---we must require $\mu = 0$.
Let us first apply an ${\rm SL}(2)$ transformation to bring the solution into the form $(a,b,c)\propto (1,0,0)$. 
This can be done without loss of generality, since we can always transform our results ex post facto to any other ${\rm SL}(2)$ frame in the full $\mu = 0$ subspace \cite{Hadar:2020kry}.
Transforming this perturbation using \eqref{transformation}
and expanding at linear order in $\lambda$, let us find how the Einstein equations constrain $\xi^\mu$.
We find that $\delta \hat E_t^{\;\;r} \propto \partial_t\xi^r - r \partial_r \partial_t \xi^r$, so $\xi^r = r[C_3(r) + C_4(t)]$ for some functions $C_{3,4}$, in terms of which $\delta\hat E_t^{\;\;t} - \delta\hat E_r^{\;\;r} \propto 2r^3 C_3'(r) + r^4 C_3''(r) + C_4''(t)$.
We thus must have $C_4''(t) = {\rm constant}$, so we write $C_4(t) = e_1 + 2e_2 t + 3 e_3 t^2$ for some constants $e_{1,2,3}$.
Solving the resulting ordinary differential equation for $C_3(r)$, we obtain the general form for $\xi^r(t,r)$,
\be
\xi^r(t,r) = e_1 r + 2e_2 r t + 3 e_3 r\left(t^2 - \frac{1}{r^2}\right) .
\ee
The final independent equation of motion is $\delta\hat E_t^{\;\;t}\propto  3 e_1 r + 6 e_2 r t + 9 e_3 r\left(t^2 + \frac{1}{r^2}\right) + 3 r \partial_t\xi^t + r^2 \partial_r \partial_t \xi^t$, which we solve to obtain the general form for $\xi^t$ parameterizing the enhanced symmetry,
\be
\xi^t(t,r) = -\left[e_0 + e_1 t + e_2 t^2 + e_3 t^3   + \frac{e_2 + 9 e_3 t}{r^2} + q_1'(r) + \frac{q_2''(t)}{r^3}\right],\label{eq:xitenhanced}
\ee
where $e_0$ is another free parameter and $q_{1,2}$ are arbitrary functions.
We can drop $q_{1,2}$, since their contribution to the perturbation will be pure gauge.
Specifically, $q_{1,2}$ can be removed by adding a pure gauge perturbation of the background, $2\lambda\nabla_{(\mu}\psi_{\nu)}$, with
\be
\begin{aligned}
\psi^t &= 2a\left[q_1(r)-rq_1'(r)+q_2(t) - \frac{q_2''(t)}{r^2} \right]\\
\psi^r &= -2a r q_2'(t).
\end{aligned}
\ee
We therefore set $q_1 = q_2 = 0$.

The accidental symmetry---the physical diffeomorphism $\xi^\mu$---can be written as
\be
\xi =  -\left[\epsilon(t)+\frac{\epsilon''(t)}{2r^{2}}+\frac{t\epsilon'''(t)}{r^{2}}\right]\partial_{t}+\left[r\epsilon'(t)-\frac{\epsilon'''(t)}{2r}\right]\partial_{r},\label{eq:xiepsilon}
\ee
where $\epsilon(t)$ is an arbitrary cubic polynomial in $t$,
\be
\epsilon(t) = e_0 + e_1 t + e_2 t^2 + e_3 t^3. 
\ee
Indeed, if we posit the form in \Eq{eq:xiepsilon}, leaving $\epsilon(t)$ arbitrary, the Einstein equations become 
\be
\epsilon''''(t) = 0. \label{eq:e4prime}
\ee
This equation has appeared in the context of JT gravity in two dimensions where, as we discuss in Sec.~\ref{Sec:Discussion}, it may be thought of as an equation of motion that puts the boundary gravitons of ${\rm AdS}_2$ on shell. 

The modes satisfying the equation of motion \eqref{eq:e4prime}, each have special physical significance.
Explicitly,
\be\label{eq:xi012}
\begin{aligned}
\xi_0  &=-(1,0)\\ \xi_1 &= -(t,-r)\\ \xi_2 &= -\left(t^2 + \frac{1}{r^2},-2 r t\right)
\end{aligned}
\ee
correspond to time translations, dilations, and special conformal transformations, respectively, while
\be\label{eq:xi3}
\xi_3 = -\left(t^3 + \frac{9t}{r^2},\frac{3}{r}-3rt^2\right)
\ee
implements the shift in the black hole's extremality.
That is, the transformations $\xi_{0,1,2}$ correspond to the infinitesimal limits of $\Delta x_{1,2,3}$, respectively, i.e., $\xi_0 = -\lim_{\alpha\rightarrow 0}\partial_\alpha \Delta x_0(\alpha)$, $\xi_1 = \lim_{\beta\rightarrow 0}\partial_\beta \Delta x_1(\beta)$, and $\xi_2 = \lim_{\gamma\rightarrow 0}\partial_\gamma \Delta x_2(\gamma)$.
The closed ${\rm SL}(2)$ algebra formed by $\xi_{0,1,2}$ is the set of isometries of the background ${\rm AdS}_2 \times S^2$, and as noted in \Sec{sec:enhancedsymmetry} we expected to find these modes.
The appearance of $\xi_3$ as a solution of \Eq{eq:eomEfinal} is nontrivial.
It represents an enhanced accidental symmetry of the Einstein equations on ${\rm AdS}_2\times S^2$, and remarkably, we find that it goes like the $t^3$ diffeomorphism akin to the on-shell boundary graviton mode in JT theory.
In terms of the ${\rm SL}(2)$ charges of the original perturbation, $(a,b,c) = (a,0,0)$, applying $\lambda \xi^\mu$ transforms the parameters as follows:
\be 
\begin{aligned}
\Delta a &=\lambda e_1 a\\
\Delta b &= 2\lambda e_2 a\\
\Delta c &= 3\lambda e_3 a.
\end{aligned}
\ee
Due to ${\rm SL}(2)$ invariance, the values of $(a,b,c)$ are not individually physically meaningful, as they can be transformed among each other.
However, in terms of the ${\rm SL}(2)$-invariant quantity $\mu = b^2 - 4 a c$, we have started with $\mu = 0$, corresponding to a perturbation that builds an extremal black hole starting from its ${\rm AdS}_2$ throat.
Under $\xi$, however, $\mu$ is modified by
\be
\Delta \mu =-4 a \Delta c = -12 \lambda e_3 a^2. 
\ee
Thus, positive $t^3$ modes in \Eq{eq:xi3} (i.e., with $\lambda e_3 < 0$) correspond to an accidental symmetry of the equations of motion relating near-horizon perturbations of an extremal RN black hole with $\mu = 0$ to those of a {\it near}-extremal hole with $\mu > 0$.

\section{Adding matter}\label{sec:mattersec}

We have shown that the linearized Einstein-Maxwell equations around the Bertotti-Robinson background possess an extra symmetry, which is manifested in the ability of the solutions to be mapped onto each other via a diffeomorphism acting only on the perturbations.
In other words,  we have investigated the source-free solutions to \Eq{eq:eomEfinal} under the action of a physicalized diffeomorphism.
In spherical symmetry, the source-free solutions that lead to an asymptotically flat RN black hole are constrained by Birkhoff's theorem so that the only nontrivial accidental symmetry present in this case was the $\xi_3$ mode, which is varying the black hole's deviation from extremality. 
This is a nonpropagating physical degree of freedom.
It is therefore worth asking whether there are physical diffeomorphisms that may also turn on propagating degrees of freedom. 
We thus investigate whether there is a larger category of perturbative equations around ${\rm AdS}_2 \times S^2$ that enjoy a similar symmetry and include a source term that supports propagating dynamics on the right-hand side of \Eq{eq:eomEfinal}, schematically,
\be 
\lim_{\lambda\rightarrow 0}\partial_\lambda {\cal E}(\bar{g}(0),h(\lambda)) = T.\label{eq:eomET}
\ee
Of course, if we allow the source $T$ to be anything at all, then certainly any transformation $\xi$ of the original electrovacuum perturbation of ${\rm AdS}_2 \times S^2$ will satisfy \Eq{eq:eomET}.
Instead, we should impose equations of motion on $T$ itself, so that it describes a self-consistent insertion of a stress-energy source.

To be concrete, let us therefore consider a source term describing a massless, minimally coupled scalar with stress-energy tensor $T_{\mu\nu}^\phi = \nabla_\mu\phi\nabla_\nu \phi - \frac{1}{2}g_{\mu\nu}\nabla_\rho \phi\nabla^\rho \phi$.
In terms of the parameter $\lambda$ describing the size of our transformation $\xi$, $\phi$ is of order $\lambda^{1/2}$.
In terms of \Eq{eq:ein} (where we input the diffeomorphism-transformed $h_{\mu\nu}$ into the perturbed metric terms and use the Bertotti-Robinson spacetime \eqref{eq:background} as background), \Eq{eq:eomET} can now be written as
\be
\delta \hat E_\mu^{\;\;\nu} = \nabla_\mu \phi \nabla^\nu\phi,\label{eq:eomETmunu}
\ee
where we require that $\phi$ satisfy the Klein-Gordon equation on ${\rm AdS}_2 \times S^2$,
\be 
\Box \phi = 0.\label{eq:KGphi}
\ee
In the s-wave sector, the most general solution for the scalar is given by \be \label{phi soln}
\phi = \sqrt{\lambda a}[f_+(v) + f_-(u)],
\ee
where $f_\pm$ are arbitrary functions of the outgoing and ingoing ${\rm AdS}_2$ null coordinates $u = t-1/r$ and $v = t+1/r$.
Equivalently, in terms of Fourier modes, the superposition of plane waves traveling up and down the black hole throat,
\be
\phi =  \sqrt{\lambda a}\left[c_+ e^{i\omega\left(t+\frac{1}{r}\right)} + c_- e^{i\omega\left(t-\frac{1}{r}\right)}\right],
\ee
satisfies \Eq{eq:KGphi} for arbitrary constants $c_\pm$.

Starting from the solution $(a,b,c) = (a,0,0)$, which describes an electrovacuum perturbation that gives rise to the empty extremal RN, we find that the stress-energy insertion \eqref{eq:eomET}, as given in \Eq{eq:eomETmunu}, may be obtained by transforming the electrovacuum solution using the transformation \eqref{transformation} with components
\be 
\begin{aligned}
\xi^t &= \phantom{+} c_+^2 e^{2i\omega\left(t+\frac{1}{r}\right)}\left[\frac{1}{16\omega^2}\left(+ 1 -\frac{2i\omega}{r} -\frac{8\omega^2}{r^2} \right) + \frac{i\omega}{4r^3} e^{-2i\omega/r} \,{\rm Ei}\left(+\frac{2i\omega}{r}\right)\right] \\ &\phantom{=} + c_-^2 e^{2i\omega\left(t-\frac{1}{r}\right)}\left[\frac{1}{16\omega^2}\left(- 1 -\frac{2i\omega}{r} +\frac{8\omega^2}{r^2} \right) + \frac{i\omega}{4r^3} e^{+2i\omega/r} \,{\rm Ei}\left(-\frac{2i\omega}{r}\right)\right] \\
&\phantom{=} + \frac{i\omega}{2r^3} e^{2i\omega t}\left[c_+ c_- \log\left(\frac{2}{r}\right) + c_0\right] \\
\\\qquad 
\xi^r &= - \frac{1}{4} \left[\left(1+\frac{ir}{2\omega}\right)c_+^2 e^{2i\omega\left(t+\frac{1}{r}\right)} +\left(1-\frac{ir}{2\omega}\right)c_-^2 e^{2i\omega\left(t-\frac{1}{r}\right)}  \right],
\end{aligned}\label{eq:xiphi}
\ee
where ${\rm Ei}(z) = -\int_{-z}^\infty \frac{e^{-x}}{x} {\rm d}x$ is the exponential integral function.
The free constant $c_0$ can be removed by adding a pure gauge perturbation $2\lambda\nabla_{(\mu}\psi_{\nu)}$ with $\psi =\lambda a c_0 e^{2i\omega t}\left[\left(\frac{2}{r^2} + \frac{1}{2\omega^2}\right)\partial_t - \frac{i r}{\omega}\partial_r \right]$.
In position space, considering the most general scalar \eqref{phi soln}, let us define two functions $F_+(v)$ and $F_-(u)$ such that $F''''_+(v) = [f'_+(v)]^2$ and $F''''_-(u) = [f'_-(u)]^2$.
Then we find that, for arbitrary $f_\pm$, we may obtain the solution to \Eq{eq:eomETmunu} by performing a physical diffeomorphism on the $(a,0,0)$ perturbation with vanishing scalar profile, using the transformation with
\be
\begin{aligned}
\xi^t =&\phantom{+}\frac{3}{2r}[F_+'(v) + F_-'(u)] - \frac{3}{2r^2}[F_+''(v)-F_-''(u)] \\
&+ \frac{3}{r^3} \left[\int^v \frac{F_+(t_0)}{(t-t_0)^4}\, {\rm d}t_0 + \int^u \frac{F_-(t_0)}{(t-t_0)^4} \,{\rm d}t_0 \right]
\\
&- \frac{1}{r^3} \int^{r} \int^t \frac{f_+'\left(\hat t + \frac{1}{\hat r}\right) f_-'\left(\hat t - \frac{1}{\hat r}\right)}{\hat r}\, {\rm d}\hat t \,{\rm d}\hat r \\
\\ \qquad 
\xi^r =&\,r[F_+'(v) - F_-'(u)] - [F_+''(v) + F_-''(u)],
\end{aligned} \label{eq:gensol}
\ee
along with adding an arbitrary superposition of $\xi_{0,1,2,3}$ as given in Eqs.~\eqref{eq:xi012} and \eqref{eq:xi3} as well as freely setting the arbitrary functions $q_{1,2}$ in $\xi^t$ as in \Eq{eq:xitenhanced}.

The existence of the solutions in Eqs.~\eqref{eq:xiphi} and \eqref{eq:gensol} is a surprising result: we have found that there exist symmetries of the equations of motion near extremal black holes---the physical diffeomorphisms $\xi$---relating perturbative solutions without matter to solutions with matter.
As a particular example, we can consider an outgoing shock wave with $T_{uu}^\phi \propto F''''_-(u) = c\, \delta(u-u_0)$.
Choosing the coefficients and functions from the constants of integration (i.e., the remaining gauge freedom) to simplify the result, the shock wave geometry is related to our original matter-free $(a,0,0)$ extremal perturbation via a physical diffeomorphism with
\be
\begin{aligned}
\xi^t &= -\frac{c}{6r^3} \left\{10 + 9 r(u_0 - t) + r^3 (u_0-t)^3 + 3 \log[r(t-u_0)]\right\}\Theta\left(t - \frac{1}{r} - u_0\right) \\
\xi^r &= \frac{c}{2r}\left[1-r^2(u_0-t)^2 \right] \Theta\left(t-\frac{1}{r}- u_0\right).
\end{aligned}
\ee

\section{Discussion}\label{Sec:Discussion}

AdS/CFT literature that considers putting large diffeomorphisms on shell classically is rather sparse. The reason for this paucity is precisely because one of the main attractions of AdS/CFT is that the gravitational theory in the bulk may be defined from an independent prescription of observables on the boundary.
That said, occasionally one does find instances in the AdS/CFT literature where certain classical actions for the large diffeomorphisms are written down and solved. 
For ${\rm AdS}_2$, the relevant context is the JT model of two-dimensional dilaton gravity, which captures many of the universal features of the spherically symmetric sector of higher-dimensional gravity near extreme black hole horizons. In JT theory, the geometry is ${\rm AdS}_2$, ${\rm d}s_2^2=-r^2 {\rm d}t^2+{\rm d}r^2/r^2$, and the dilaton is identified with $\Phi_{\rm JT}=\Phi= a r + b r t + c r(t^2 - r^{-2})$ from Eq.~\eqref{Phi}. The large diffeomorphisms of ${\rm AdS}_2$, in FG gauge, are given by \cite{Mandal:2017thl,Gaikwad:2018dfc}
\begin{align}\label{AdS2 large diffeo}
&t\to f(t)+\frac{2f''(t)f'(t)^2}{4r^2f'(t)^2-f''(t)^2},\qquad r\to \frac{4r^2f'(t)^2-f''(t)^2}{4rf'(t)^3},
\end{align}
under which the metric and dilaton become
\begin{align}\label{AdS2 transformed}
&{\rm d}s_2^2\to -r^2\left(1+\frac{{\rm Sch}(f,t)}{2r^2}\right)^2{\rm d}t^2 +\frac{{\rm d}r^2}{r^2}\qquad {\rm and}\qquad \Phi\to \phi_0(t)r+\frac{v(t)}{r},
\end{align}
with the Schwarzian derivative defined by ${\rm Sch}(f,t) =(f''/f')'-\frac{1}{2}(f'')^2/(f')^2$ and the source and vev in $\Phi$ being $\phi_0(t)=[a+b f(t)+c f(t)^2]/f'(t)$ and $v(t)=-[\phi_0''(t)+{\rm Sch}(f,t)\phi_0(t)]/2$, respectively.
For arbitrary $f$, this source satisfies the equation of motion
\be\label{Schwarzian eom}
\left[\frac{1}{f'}\left(\frac{(f' \phi_0)'}{f'}\right)'\right]'=0,
\ee
which one may also obtain by varying the action $I=\int {\rm d}t \,\phi_0(t)\, {\rm Sch}(f,t)$ with respect to $f$. 
Notice that in standard AdS/CFT spirit, the above equation of motion would be viewed simply as an equation that for given arbitrary $f(t)$ prescribes a boundary source $\phi_0(t)$ as dictated by integration of this equation.
On the other hand, this equation constrains the allowed $f(t)$ whenever one independently chooses the value of the source $\phi_0(t)$. For example, if one imposes that $\phi_0(t)={\rm constant}$, as discussed, e.g., in Ref.~\cite{Nayak:2018qej}, then for infinitesimal diffeomorphisms $f(t)=t+\epsilon(t)$, Eq.~\eqref{Schwarzian eom} reduces to $\epsilon''''(t)=0$ as in \Eq{eq:e4prime}. 
Therefore imposing $\phi_0(t)={\rm constant}$ before as well as after acting with the diffeomorphism \eqref{AdS2 large diffeo}, which is to say imposing that the first term in Eq.~\eqref{eq:eomElinear} vanishes, is akin to requiring that the large diffeomorphism is such that, when a connection with an asymptotically flat region is maintained, the Einstein equations are satisfied across the entire spacetime. The analogy with our analysis in Secs.~\ref{sec:symsec} and \ref{sec:mattersec} continues to hold in the case when matter is added, too, with Eq.~\eqref{Schwarzian eom} acquiring a corresponding matter source in the right-hand side as well. In the case of ${\rm AdS}_3$, the Schwarzian action $I=\int {\rm d}t\, {\rm Sch}(f,t)$ that puts large ${\rm AdS}_2$ diffeomorphisms on shell is replaced by a Liouville action for large ${\rm AdS}_3$ diffeomorphisms \cite{Carlip:2005tz} (see also Ref.~\cite{Cotler:2018zff}).

It is interesting to consider what the precise off-shell extension of our transformations could be and compare it with existing literature on ${\rm AdS}_2$ asymptotic symmetry group generators, such as the ones found in Refs.~\cite{Cadoni:1999ja,Navarro-Salas:1999zer}. In the electrovacuum case, the above discussion implies that the mode \eqref{eq:xi3} is to be identified with the $L_{-2}$ mode of a Virasoro algebra whose ${\rm SL}(2)$ elements $\{L_0,L_{\pm 1}\}$ are given by Eq.~\eqref{eq:xi012}. Then we can generate all the Virasoro modes by computing Lie brackets starting from Eqs.~\eqref{eq:xi012} and \eqref{eq:xi3}. Alternatively, we may guess how to generalize Eq.~\eqref{eq:xiepsilon} to arbitrary $\epsilon(t)=t^n$, without impacting the modes $\xi_{0,1,2,3}$, such that we get Virasoro-generating vector fields $\hat{\xi}$ up to terms of ${\cal O}(1/r^3)$. To this end, we can add to Eq.~\eqref{eq:xiepsilon} any combination $\propto \epsilon''''(t)$; we find that upon writing
\be
\hat\xi = -\left[\epsilon(t)+\frac{\epsilon''(t)}{2r^{2}}+\frac{t\epsilon'''(t)}{r^{2}}-\frac{t\epsilon''(t)\epsilon''''(t)}{r^{2}\epsilon'''(t)}\right]\partial_{t}+\left[r\epsilon'(t)-\frac{\epsilon'''(t)}{2r}\right]\partial_{r},\label{eq:xihat}
\ee
which leaves Eqs.~\eqref{eq:xi012} and \eqref{eq:xi3} unchanged, the modes $\hat{\xi}_n$ for $\epsilon(t)=t^n$ satisfy a Virasoro algebra,
\be 
[\hat\xi_m,\hat\xi_n] = (m-n)\hat\xi_{m+n-1},\label{eq:Virasoro}
\ee
up to ${\cal O}(1/r^3)$. The leading terms in $\hat{\xi}$ agree with the ones found in Refs.~\cite{Cadoni:1999ja,Navarro-Salas:1999zer}, and it would be interesting to know whether there exist ${\rm AdS}_2$ boundary conditions that produce the subleading terms in Eq.~\eqref{eq:xihat} as well.

Prior to the recent surge of JT gravity studies, another large body of work where the asymptotic symmetries of ${\rm AdS}_2$ have been shown to contribute to physical quantities associated with the near-horizon of extreme black holes may be found in calculations of the leading (logarithmic) corrections to the Bekenstein-Hawking area-entropy law by Sen et al. \cite{Banerjee:2010qc, Banerjee:2011jp, Sen:2012kpz,Sen:2012cj, Bhattacharyya:2012wz, Sen:2012dw}. In this earlier work, the large diffeomorphisms of ${\rm AdS}_2$ contribute, as zero-modes, to the partition function that computes the black hole entropy. 
It is interesting to contemplate whether an off-shell extension of our transformations, such as Eq.~\eqref{eq:xihat}, may be associated to the specific zero-modes that contribute to the logarithmic correction of black hole entropy calculations. 
In this respect, there are at least two reasons for hesitation regarding the existence of such an association. First, in our analysis we consider the action of our transformations on physical perturbations that change the size of the $S^2$ transverse to ${\rm AdS}_2$, whereas in the work of Sen et al. the transverse sphere is held fixed (it is precisely in the context of JT gravity that this has been lifted later). Second, as noted in Sec.~\ref{sec:enhancedsymmetry}, our transformations act only on the physical perturbative solutions around the background ${\rm AdS}_2\times S^2$ and not on the background itself, which is again in contrast with the zero-mode contributions to black hole entropy corrections, which come entirely from the action of large diffeomorphisms on the exact ${\rm AdS}_2\times S^2$ throat geometry.

${\rm AdS}_2$-like geometries arise universally in the near-horizon limit of black holes that are close to extremality, and we expect the main results of this paper to extend more broadly to all such near-horizon geometries. Within the context of spherical symmetry, it is interesting to investigate what new features might arise from nonminimal coupling of propagating scalar degrees of freedom on an ${\rm AdS}_2\times S^2$, such as the one recently identified in Ref.~\cite{Porfyriadis:2021zfb}. Similarly, since it is known that for rotating ${\rm AdS}_2$-like near-horizon solutions the universal aspects of JT gravity are complemented by additional new features \cite{Castro:2018ffi, Moitra:2019bub, Castro:2021csm, Castro:2021fhc}, it is of great interest to investigate our accidental symmetries for axisymmetric propagating gravitational modes in that context as well.
We leave open avenues of investigation to future work.

\vspace{5mm} 
 
\begin{center} 
{\bf Acknowledgments}
\end{center}
\noindent 
We thank Shahar Hadar, Gary Horowitz, Adam Levine, Alex Lupsasca, Andy Strominger, and Ying Zhao for useful discussions and comments.
A.P.P. is supported by the Black Hole Initiative at Harvard University, which is funded by grants from the John Templeton Foundation and the Gordon and Betty Moore Foundation.
G.N.R. is supported at the Kavli Institute for Theoretical Physics by the Simons Foundation (Grant~No.~216179) and the National Science Foundation (Grant~No.~NSF PHY-1748958) and at the University of California, Santa Barbara by the Fundamental Physics Fellowship.

\vspace{3mm}

\bibliographystyle{utphys-modified}
\bibliography{Diffeos}

\end{document}